\documentclass[prl,twocolumn,10pt,aps,longbibliography,superscriptaddress]{revtex4-1}
\usepackage{epsfig,amsmath,amssymb,amsfonts,graphicx,color,calc,epstopdf}
\usepackage{xspace} 
   \usepackage[T1]{fontenc}

   \let\d=\delta
   
\let\l=\lambda

\let\D=\Delta   
   
 \let\r=\rho

\def\DD{{\cal D}}
\def\GG{{\cal G}} \def\SS{{\cal S}}
  
\def\ZZ{{\cal Z}}

\def\to{\rightarrow}

\def\de{\mathrm d}

\newcommand{\beq}{\begin{equation}} 
\newcommand{\eeq}{\end{equation}}
\newcommand{\ba}{\begin{eqnarray}}
\newcommand{\ea}{\end{eqnarray}}

\def\de{\mathrm d}

\begin{document}

\title{
On the glassy nature of the hard phase in inference problems
}

 \author{Fabrizio Antenucci}
\affiliation{
Institut de physique th\'eorique, Universit\'e Paris Saclay, CNRS, CEA, F-91191 Gif-sur-Yvette, France
}
\affiliation{Soft and Living Matter Lab., Rome Unit of CNR-NANOTEC, Institute of
Nanotechnology, Piazzale Aldo Moro 5, I-00185, Rome, Italy}

\author{Silvio Franz}
\affiliation{LPTMS, CNRS, Univ. Paris-Sud, Universit\'e Paris-Saclay, 91405 Orsay, France}

\author{Pierfrancesco Urbani}
\affiliation{
Institut de physique th\'eorique, Universit\'e Paris Saclay, CNRS, CEA, F-91191 Gif-sur-Yvette, France
}

\author{Lenka Zdeborov\'a}
\affiliation{
Institut de physique th\'eorique, Universit\'e Paris Saclay, CNRS, CEA, F-91191 Gif-sur-Yvette, France
}

\begin{abstract}
An algorithmically hard phase was described in a range of 
inference problems: even if the signal can be reconstructed with a
small error from an information theoretic point of view, known algorithms
fail unless the noise-to-signal ratio is sufficiently small.
This \emph{hard phase} is typically understood as a metastable branch
of the dynamical evolution of message passing algorithms.
In this work we study the metastable branch for a prototypical inference problem, the low-rank
matrix factorization, that presents a hard phase.
We show that for noise-to-signal ratios that are below the information
theoretic threshold, the posterior measure is composed of an
exponential number of metastable glassy states and we compute their entropy, called the complexity.
We show that this glassiness extends even slightly below the
algorithmic threshold below which the well-known approximate message
passing (AMP)
algorithm is able to closely reconstruct the signal. 
Counter-intuitively, we find that the performance of the AMP
algorithm is not improved by taking into account the glassy nature
of the hard phase. 
 This provides further evidence that the hard phase in
  inference problems is algorithmically impenetrable for some deep
  computational reasons that remain to be uncovered.

\end{abstract}

\maketitle

\section{Introduction} 
Inference problems are ubiquitous in many scientific areas involving data. 
They can be summarized as follows: a signal is
measured or observed in some way and the inference task is to reconstruct
the signal from the set of observations. Many practical applications
involving data  rely on our ability to solve inference problems fast and efficiently. While from the point of view of
computational complexity theory many of the practically important
inference problems are algorithmically hard in the worst case,
practitioners are solving them every day in many cases of interest. It is
hence an important research question to know which types of inference
problems can be solved efficiently and which cannot. Formally
satisfying answer to this question would lead to an entirely new
theory of typical computational complexity, and would likely shed new
light on the way we develop algorithms. 

For a range of inference problems the Bayesian inference naturally leads to statistical physics
of systems with disorder, see e.g. \cite{grassbergerstatistical}. This
connection was explored in a range of recent works and brought a class
of models for inference problem in which the Bayes-optimal inference
can be analyzed and presents a first order phase transition. As common
in physics in high dimension, the first
order phase transition is associated to the existence of a
metastable region in which known efficient algorithms fail to reach
the theoretical optimal performance. This metastable region was coined as the {\it
  hard phase}, see e.g. \cite{KZ16}. It has been located in error correcting codes \cite{richardson2008modern,MM09},
compressed sensing \cite{krzakala2012statistical}, community detection
\cite{decelle2011asymptotic}, the hidden-dense submatrix problem
\cite{deshpande2015finding,montanari2015finding}, low-rank
estimation problems including data clustering, sparse PCA or tensor
factorization \cite{richard2014statistical,LMLKZ17}, learning in neural
networks \cite{gyorgyi1990first}.
The nature of the hard phase in all these problems is of the same
origin, and therefore it is expected that algorithmic improvement in
any of them would lead to improvement in all the others as well.

In the current state-of-the-art (including the references above) the hard phase is located as a
performance barrier of a class of message passing algorithms. 
Message passing algorithms can be seen as spin-offs of the cavity
method of spin glasses \cite{MPV87}. In the context of inference on
dense graphical models the algorithms is called approximate
message passing (AMP) known from the context of compressed sensing
\cite{DMM09}.
In the limit of large system size, the dynamical evolution of AMP
can be tracked by the so-called \emph{state evolution} (SE) \cite{DMM09,bayati2011dynamics}, whose fixed point
equations coincide with the saddle point equations describing the
thermodynamic of the system under the {\it replica symmetric} 
assumption.
The analysis of SE and its comparison to the analysis of the
Bayes-optimal performance reveals that there is an interval of
noise-to-signal ratio where the signal could be
reconstructed by sampling the posterior measure,  while
AMP is not able to converge to the optimal error. This interval marks
the presence of the \emph{hard phase}.

In this paper we want to attract further attention of the physics
community towards the existence of this hard phase related to a 1st
order phase transition in the optimal performance in inference
problems. The following open questions might use the physics-like
approach and insights: Could there be a physics-inspired algorithm
that is able to overcome the algorithmic barrier the AMP algorithm
encounters? Note that in problems where the corresponding graphical
model can be designed, such as compressed sensing or error correcting
codes, such a
strategy related to nucleation indeed exists \cite{kudekar2011threshold,krzakala2012statistical}. But what about the
more ubiquitous problems where the graphical model is fixed? Are there
some physical principles or laws that can provide further evidence
towards the impenetrability of the algorithmic barrier? 

The motivation of the present work was to investigate the above
questions.
We analyze the following physics-motivated strategy: It is known that the
metastable part
of the posterior measure in the hard phase is glassy
\cite{sompolinsky1990learning,0295-5075-55-4-465,krzakala2009hiding}.
Yet, the AMP algorithm  fails to describe this glassiness properly. 
In some other contexts where 
message passing algorithms
are successfully used, a correct account of glassiness leads to algorithm
that improve over simpler ones. Notably this is the case of
random constraint satisfaction problems, where the influential work
\cite{MPZ02} has shown that {\it survey propagation}, that takes
correctly glassiness into account, beats the performance of {\it belief propagation}. 

We pose therefore the problem whether, in inference tasks, the
reconstruction of the signal becomes easier when one uses algorithms
in which the glassiness is correctly 
taken into account.  We investigate this strategy thoroughly in the present work. 
We confirm that the hard phase is glassy in the sense that it consists of
an exponential number of local optima at higher free energy than the
equilibrium one. However, when it comes to the 
reconstruction of the signal, our analysis leads us to the remarkable
conclusion that, in contrast to constraint satisfaction and optimization
problems, in inference problems taking into account the glassiness of
the hard phase does not improve upon the performance of the simplest AMP
algorithm. 
We thus provide an additional evidence towards the bold
conjecture that in the corresponding inference problems AMP is the
best of low-computational-complexity inference algorithms.

Note that such a {\it negative} result is very interesting from both
physics and computer science point of view. In physics, a common intuitive narrative
tells us that the properties of the energy landscape control the
algorithmic difficulty of the problem. Yet a solid and physically
intuitive explanation of why inference algorithm could not penetrate
the hard phase remains open. Our results invite researchers to progress
in this question, eventually leading to a precise understanding
of the interplay between dynamics and landscape. In computer science, developments that 
go beyond the traditional worst-case computational complexity
results are rare and the hard phase provides an unique and a sharply
delimited case that might be computationally hard even for a typical
instance. Building a theory that would explain the nature of hard
phase might be the next pillar of our understanding of computational complexity.

Our analysis of the glassiness of the hard phase provides new insights
on the performance of Monte Carlo or Langevin dynamics. Presence of
the glassiness suggests that these sampling-based algorithms
are slowed-down and thus their commonly used versions may not be able
to match the performance of AMP. While this aligns with some of the
the early literature \cite{sompolinsky1990learning}, more recent
literature \cite{decelle2011asymptotic} suggested, based on numerical
evidence, that Monte Carlo
sampling is as good as the message passing algorithm. Based on
conclusion of our work, this question of performance barriers of
sampling-based algorithms should be re-opened and investigated more
thoroughly. Good understanding of performance of these algorithms is
especially important in the view of the fact that some of the most
performing systems currently use stochastic gradient descent, that can
be seen as a variant of the Langevin dynamics. 

This paper is organized as follows. In Section \ref{sec:model} we
introduce the model on which we illustrate the main findings of this
paper, we expect this picture to be generic and apply to all the
models where the hard phase related to a first order phase transition
in the performance of the Bayesian inference was identified. In
Section \ref{sec:Bayes} we remind the basic setting of Bayesian
inference.  In Section \ref{sec:main_results} we give a summary of the main algorithmic
consequences of our work. In Section \ref{sec:replicas} we then remind the replica
approach to the study of the corresponding posterior measure.
Section \ref{sec:RS} then summarized the known replica symmetric
diagram and the resulting phase transitions. Section \ref{sec:1RSB}
then includes the main technical results of the paper where we
quantitatively analyze the glassiness of the hard phase, giving rise
to 
our conclusions in section \ref{sec:conclusion}.  

\section{Model} 
\label{sec:model}
In order to be concrete we concentrate on a prototypical example of an
inference problem with a hard phase - the constrained rank-one matrix
estimation. This problem is representative of the whole
class of inference problems where the hard phase related to a 1st
order phase transition was identified \cite{deshpande2015finding,lesieur2015phase,LKZ17}. We choose this example because it
is very close to the Sherrington-Kirkpatrick model for which the study
of glassy states is the most advanced \cite{MPV87}. Glassiness was also studied in
detail in the spherical or Ising $p$-spin model, corresponding to
spiked tensor estimation \cite{richard2014statistical}. However, in that model the hard phase spans the
full low-noise phase and the transition towards the easy phase, on
which we aim focus here, happens for noise-to-signal-ratio too low to be
straightforwardly investigated within the replica method. 

In the rank-one matrix estimation problem the signal, denoted by $\underline x^{(0)}\in {\mathbb R}^N$, 
is extracted from some separable prior probability distribution given by $\underline P_X(\underline x^{(0)}) = \prod_{i=1}^N P(x^{(0)}_i)$.
This signal is subjected to noisy measurements of the following form
\beq
\begin{split}
Y_{ij} &= \frac{1}{\sqrt N} \, x_{i}^{(0)} x_{j}^{(0)} + \xi_{ij} \, , \ \ \ \ \forall\ i\leq j
\end{split}
\label{rankone}
\eeq
where $\xi_{ij}$ are Gaussian random variables with zero mean and variance $\Delta$. 
Therefore one observes the signal through the matrix~$Y$.
The inference problem is to reconstruct the signal $\underline x^{(0)}$ given the observation of the matrix $Y$.
The informational-theoretically optimal
performance in this problem was analyzed in detail in \cite{LKZ17}
and this analysis was proven rigorously to be correct in
\cite{deshpande2014information,krzakala2016mutual,barbier2016mutual,lelarge2016fundamental}. 
Refs. \cite{rangan2012iterative,deshpande2014information,LKZ17} also
analyzed the performance of the AMP algorithm. 

While the theoretical part of this paper is for a generic prior $P_X$,
the results section focuses on the Rademacher-Bernoulli prior 
\beq
P_X(x) = \left(1-\r\right)\d(x) + \frac{\r}{2}\left[\d(x-1)+\d(x+1)\right]
\label{eq:RadBern}
\eeq
as this is a prototypical yet simple example in which the hard phase appears for
sufficiently low $\rho$ \cite{lesieur2015phase,LKZ17}. Let us mention that the rank-one
matrix estimation with the Rademacher-Bernoulli prior has a very
natural interpretation in terms of community detection
problem. Keeping this interpretation in mind can help the reader to
get intuition about the problem. Nodes
are of three types: $x^{(0)}=1$ belong to one community, $x^{(0)}=-1$ to a second
community, and $x^{(0)}=0$ does not belong to any community. The
observations $Y_{ij}$ (\ref{rankone}) can be interpreted as 
weights on edges of a graph that are on
average larger for nodes that are either both in community one or both
in community two, they are on average smaller if one of the nodes is
in community one and the other in community two, and they are
independent and unbiased when one of the nodes does not belong to any
community. Thanks to the output universality result of \cite{lesieur2015mmse,krzakala2016mutual} the
result presented in this paper also hold for a model where the
observations $Y_{ij} \in \{0,1\}$ correspond to the adjacency matrix
of an unweighted graph with Fisher information corresponding to the
inverse of the variance $\Delta$.   

\section{Bayesian inference and approximate message passing} 
\label{sec:Bayes}
We study the the so-called Bayes optimal setting, which means that we
know both the prior $\underline P_X(\underline x)$ and the variance
$\D$ of the noise. The probability distribution of $\underline x$ given $Y$
is given by Bayes formula
\beq
P(\underline x| Y) \propto P_X(\underline x) P(Y|\underline x)\:.
\label{Bayes}
\eeq
Since the noise $\xi_{ij}$ is Gaussian we have
\beq
\begin{split}
P(Y|\underline x) &\propto \prod_{i\leq j} \exp\left[-\frac 1{2\D} \left(Y_{ij}-\frac{x_ix_j}{\sqrt N}\right)^2\right]\\
&\equiv \prod_{i\leq j} \GG\left(Y_{ij}|\frac{x_ix_j}{\sqrt N}\right)\:.
\label{output}
\end{split}
\eeq
Both in Eq.~(\ref{Bayes}) and (\ref{output}) we have omitted the normalization constants.
An estimate of the components of the signal that minimize the
mean-squared-error with the ground truth signal $\underline x^{(0)}$ is
computed as
\beq
\hat x_i = \langle x_i\rangle
\eeq
where the brackets stand for the average over the posterior measure Eq.~(\ref{Bayes}).
Therefore in order to solve the inference problem we need to compute
the local magnetizations $\{\hat x_i\}$.
The AMP algorithm is aiming to do precisely that, its derivation can
be found e.g. in \cite{LKZ17}.
AMP boils down to a set of recursion relations of the form \beq \hat
x_i^{(t+1)} = \textrm{AMP}_i\left(\underline{\hat x}^{(t)}, {\hat
  x}^{(t-1)}_i\right) \, , \label{eq:AMP}\eeq whose iterative fixed point is taken as an
estimate of the signal.  
It is known that fixed points of the state evolution of the AMP
algorithm is in the
thermodynamic limit described by the replica symmetric (RS)
solution of the model \cite{DMM09,bayati2011dynamics}. AMP follows the RS solution irrespectively of
the fact whether RS is the physically correct description of the posterior measure or not. 

As shown in \cite{AKUZ18}, it is possible to
derive a generalized AMP, that we call \textit{Approximate Survey
  Propagation } (ASP) algorithm, whose
state evolution fixed points coincide with the replica equations in
the one-step replica symmetry breaking (1RSB) ansatz. Just as AMP, the
ASP algorithm can be also written in a form \cite{AKUZ18} 
\beq \hat x_i^{(t+1)} = \textrm{ASP}_i\left(\underline{\hat x}^{(t)}, {\hat
  x}^{(t-1)}_i,s\right) \, ,  \label{eq:ASP}\eeq
depending on one additional free parameter $s$, corresponding to the Parisi
parameter from the spin glass literature. The special case of $s=1$
reduces the ASP algorithm back to AMP. The 1RSB solution is known to
provide a better description - in many case exact - of glassy states.  
In section \ref{sec:replicas} we hence study the
thermodynamics of the above model in the RS
and 1RSB ansatz, focusing on its properties in the hard phase.

\section{Summary of main algorithmic result} 
\label{sec:main_results}

Before going to the technical part of the replica analysis in
Sec.~\ref{sec:replicas}, we briefly summarize the corresponding main
algorithmic result. In section \ref{sec:replicas} we then investigate in detail
the 1RSB solution of the low-rank matrix estimation model
(\ref{rankone}) focusing on the glassy properties of the hard
phase. Our main interest, however, is in the relation between the 1RSB
solution and the associated algorithmic
performance. The main question we ask is
whether ASP can (for a suitable choice of the Parisi
parameter $s$) improve on AMP. The experience with
survey propagation algorithm applied to constraint satisfaction
problems \cite{BMZ05} suggests that this should be possible. 

In Fig.~\ref{fig:PD} we plot the magnetization achieved by the ASP algorithm as a
function of the noise $\Delta$ for several values of the Parisi
parameter s. We observe that as the noise $\Delta$ decreases the
equilibrium value (yellow) is reached first by the $s=1$ curve, corresponding to performance of 
AMP. In Fig.~\ref{fig:MSE} we then
plot the mean-squared-error as a function of the Parisi parameter $s$
for several values of the noise $\Delta$. Again we see that in all
cases the best error is achieved with $s=1$. Algorithmically this means
that in the present setting, ASP never obtains better accuracy than
the canonical AMP algorithm.

The fact that among all the values of $s$ the lowest MSE is reached by the $s=1$ states for all
$\Delta$ is unexpected from the physics point of view. It implies that the AMP that neglects glassiness and 
wrongly describes the hard region works better as an inference algorithm 
than an algorithm that correctly describes the metastable
states in this region. 
At the same time, the above result could be anticipated based on
mathematical theorem of \cite{deshpande2015finding} that implies
that AMP is optimal among all local algorithms. This theorem applies as long as an iterative algorithm only uses
information from nearest neighbours and (nearly) reaches a fixed
point after $O(1)$ iterations. 

\section{The replica approach to the posterior measure}
\label{sec:replicas}
In order to study the posterior measure, we define the corresponding free energy as
\beq
\mathrm f\left[\D; Y\right] =-\frac 1N \ln \int \left(\prod_{i=1}^N
  \de x_i P_X(x_i)\right) \prod_{i\leq j} \GG\left(Y_{ij} \bigg|
  \frac{x_ix_j}{\sqrt N}\right)\, .
\eeq
This is a random object since it depends on the matrix~$Y$. Furthermore it depends on $\D$ through the function~$\GG$. 
Indeed, we want to study the typical behavior of this sample-dependent free energy.
Therefore we define
\beq
\mathrm f(\D)=\overline{\mathrm f\left[\D; Y\right]} \equiv \int
\left[\prod_{i\leq j} \de Y_{ij} \right]P(Y) \mathrm f\left[\D;
  Y\right] \, ,
\label{av_s}
\eeq where $Y$ is obtained as in Eq.~(\ref{rankone}), so that $P(Y)$
is given by \beq P(Y)\propto \int \de \underline x^{(0)}\, \underline
P_X\left(\underline x^{(0)} \right) \prod_{i\leq j}
\GG\left(Y_{ij}\bigg|\frac{x_i^{(0)}x_j^{(0)}}{\sqrt N}\right)\:.  \eeq In
order to perform the average defined in Eq.~(\ref{av_s}) we use the
replica method \cite{MPV87}.  Introducing \beq \ZZ = \int
\left(\prod_{i=1}^N \de x_i P_X(x_i)\right) \prod_{i\leq j}
\GG\left(Y_{ij}\bigg|\frac{x_ix_j}{\sqrt N}\right)\, , \eeq we get \beq \mathrm
f(\D) =- \frac 1N\lim_{n\to 0} \partial_n \int \left[\prod_{i\leq j}
  \de Y_{ij} \right]P(Y) \ZZ^n \, .\eeq For integer $n$ we can represent
$\ZZ^n$ as an $n$-dimensional integral over $n$ replicas $\underline
x^{(a)}$ with $a=1,\ldots, n$. Stated in this way the problem is
obviously symmetric under the exchange of the $n$ replicas among
themselves. Moreover since we need to integrate over the signal distribution
$P(Y)$ we end up with a system of $n+1$ replicas,
that, in the Bayes optimal
case, is symmetric under the
permutation among \emph{all} the
$n+1$ replicas.  Performing standard manipulations, see e.g. \cite{MPV87}, we
arrive at a closed expression for $\mathrm f (\D)$ that is \beq
\mathrm f(\D) = -\frac{1}{N}\ln \int \DD q \DD \hat q \exp\left[N
  \SS\left(q, \hat q\right)\right]\, ,
\label{replicated_f}
\eeq 
where $\SS$ is a function that can be computed explicitly and $q$ and $\hat q$ are $(n+1)\times (n+1)$ overlap matrices.
In the large $N$ limit, the integral in Eq.~(\ref{replicated_f}) can
be evaluated using the saddle point method.  At the saddle point level
the physical meaning of the overlap matrix $q$ is given in terms of
\beq q_{ab} = \frac 1N \sum_{i=1}^N \overline{\left\langle x_i^{(a)}
  x_i^{(b)}\right\rangle } \, ,\eeq while the matrix $\hat q$ is just a
Lagrange multiplier.  We denote $m$ the magnetization of the system,
meaning \beq m\equiv q_{0a} = q_{a0} = \frac 1N \sum_{i=1}^N
\overline{\left\langle x_i^{(0)}x_i^{(a)} \right\rangle} \ \ \ a>0\:.
\eeq The saddle point equations for $q$ and $\hat q$ can be written in
complete generality for any $n$ but then one needs to take the
analytic continuation down to $n\to 0$.  One needs an
appropriate scheme from which one can take the replica limit.  Here we
consider two schemes: the replica symmetric (RS) and
the 1-step replica symmetry breaking (1RSB) one.
We refer here to
symmetry under permutations of the $n$ replicas with index $a=1,\dots,n$.

 \subsection{Reminder of the replica symmetric solution} 
 The RS scheme boils down to consider \beq
\begin{split}
q_{ab}&=\left(q_d - q_0\right)\d_{ab} + q_0 \ \ \ \ a,b\geq 1\, ,\\
\hat q_{ab}&=\left(\hat q_d - \hat q_0\right)\d_{ab} + \hat q_0 \ \ \
\  a,b\geq 1\, ,\\
q_{0a} &= q_{a0} = m \ \ \ \ a\geq 1 \, ,\\
\hat q_{0a} &= \hat q_{a0} = \hat m \ \ \ \ a\geq 1\, .
\label{eq:RS:ansatz}
\end{split}
\eeq\label{sec:RS} 

From the point of view of the inference, the relevant quantity to look at is the Mean Square Error (MSE)
\beq
\begin{split}
 \text{MSE} =&  \frac 1N \sum_{i=1}^N \overline{ \left( \langle x_i\rangle - x_i^{(0)} \right)^2 }
 \\
 =&  \rho - 2 m + q_0\, ,
 \label{eq:MSE}
\end{split}
\eeq
where $\rho \equiv \overline{ \left\langle x^{(0)} \right\rangle^2}$.
Replica symmetry among all the $n+1$ replicas is obtained for
$m=q_0$. It is well known that, as a direct
consequence of Bayes optimality (also called Nishimori condition
\cite{KZ16}), this fully replica symmetric solution
is the one that describes thermodynamically dominant states. The more
general ansatz is, however, important as it allows to 
describes metastable states where the Nishimori identities might not hold.  
Plugging this ansatz inside the expression for $\SS$
and taking the saddle point equations w.r.t. all these parameters one
gets the replica symmetric solution as reported in \cite{LKZ17}, and
proven to give the equilibrium solution in
\cite{barbier2016mutual,lelarge2016fundamental}.  The RS free energy
can be expressed as 
\begin{align}
 \mathrm f_{\text{RS}} (\D) =  \min_{m}\left\{ \phi_{\text{RS}} \left( m, \D \right) 
 \right\}
\end{align}
with
\begin{align}
 \phi_{\text{RS}} \left( m , \D \right) =  \frac{m^2}{4 \D}  - \mathbb{E}_{x^{(0)},W} \left[ f \left( \frac{m}{\D} , \frac{m}{\D} x^{(0)} + \sqrt{\frac{m}{\D}} W \right) \right]
 \label{eq:RS_FE}
\end{align}
where
\begin{align}
 f \left( A, B \right) =
 \ln \left[ \int dx \, P_X (x) \, e^{ - \frac{1}{2} A x^2 + B x}
  \right]\, ,
\end{align}
and $x^{(0)}$ and $W$ are random variables distributed
according $P_X \left( x^{(0)} \right)$ and a standard
normal distribution, respectively.
The values of $m$ for which $\phi_{\text{RS}}$ is stationary are the solution of
\begin{align}
 m = \mathbb{E}_{x^{(0)},W} \left[ x^{(0)} \frac{\partial f}{\partial B}  \left( \frac{m}{\D} , \frac{m}{\D} x^{(0)} + \sqrt{\frac{m}{\D}} W \right) \right] \, .
\end{align}
Equilibrium properties of the inference problem are given by the global
minima of
the free energy Eq.~(\ref{eq:RS_FE}). Local minima of the free
energy that do not correspond to the
equilibrium solution are called {\it metastable}.

\begin{figure}[t]
\includegraphics[width=\columnwidth]{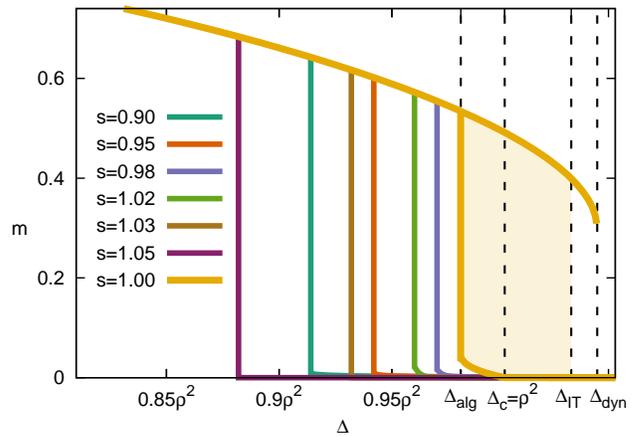}
\caption{The magnetization, \emph{aka} the overlap, between the signal
  and the states described by the 1RSB solution at Parisi parameter
  $s$, as a function of the noise strength $\D$, and sparsity $\rho=0.08$. The curve
  that show a spinodal transition towards the strongly magnetized
  solution at largest values of $\D$ is the one for $s=1$. The same curve represents also the
  performance of the AMP algorithm. Taking the glassiness of the
  metastable branch into account does not improve upon AMP. }
\label{fig:PD}
\end{figure}

For illustration, we consider the case of the Rademacher-Bernoulli
prior (\ref{eq:RadBern}) and we set $\r=0.08$ so that the inference problem has an hard phase \cite{LKZ17}.
The replica symmetric phase diagram is represented in Fig.~\ref{fig:PD} (yellow curve).

At high $\D$ the noise is so strong that the signal cannot be
recovered and therefore $m=0$. Upon decreasing $\D$ the signal is
relatively stronger w.r.t the noise and for $\D = \D_{\rm dyn} \sim 1.041 \rho^2$ the
system undergoes a {\it dynamical transition}. On the one hand one can
see that the free energy (\ref{eq:RS_FE}) develops a local metastable
 minimum with $m>0$. On the other hand, the $m=0$ state undergoes a
clustering transition according to the pattern familiar in
the physics of spin glasses \cite{FP95,CC05}. The corresponding
RS free energy ceases to describe a paramagnetic state and it
describes a non-ergodic phase with an exponential number
$\exp(N\Sigma(\D))$ of metastable states - aka clusters - with zero
overlap among each other and identical energy and internal
entropy. Both the zero $m$ dominating branch and the metastable $m>0$
branch have identical energy and internal entropy. Their free energy
difference is the complexity $f(m>0)-f(m=0)=\Sigma(\D)$. Moreover, as
we will see in the next section, the typical overlap $q_1$ between
configurations in these states coincides with the value of $m$ of the magnetized solution.  
For that reason the magnetized
state corresponds just to one cluster among the exponential
multiplicity dominating the thermodynamics. The complexity (i.e. log
of their number) of the
thermodynamic states decreases with~$\D$, until it vanishes at a value
$\D=\D_{\rm IT} \sim 1.0295 \rho^2$ where there is the information theoretic phase
transition and $\Sigma(\D_{\textrm{IT}})=0$. The signal is here strong enough so that a first order
phase transition happens where the minimum with positive magnetization
becomes the global minimum of the free energy. The complexity of the
$m=0$ solution becomes negative, the solution is non
physical and consequently RSB is necessary to describe the metastable branch. Despite this fact, 
this RS metastable branch cannot be just dismissed as unphysical: it
continues to be relevant algorithmically as dynamical attractor of the
AMP algorithm.  Decreasing the intensity of the noise further, another phase
transition happens in this RS  branch. At $\D=\D_c = \rho^2$ the metastable minimum
develops a small magnetization. Decreasing even further $\D$, at
$\D = \D_{\textrm{alg}} \sim 0.9805 \rho^2$ this metastable minimum disappears with a spinodal
transition. In the  interval
$[\D_{\textrm{alg}},\Delta_{\rm IT}]$ one finds the hard phase defined
by the property that the AMP algorithm is sub-optimal
 (the shaded yellow region in Fig.~\ref{fig:PD}): the
global minimum of the free energy has a high $m$ (low MSE), but the
small $m$ non-physical local minimum  continues to describe the
attractor of the AMP. The state evolution describing the AMP algorithm starting
from random conditions converges to the local minimum of
lowest magnetization.

\subsection{Glassy phase and complexity}
\label{sec:1RSB}

The low branch RS solution is non-physical below $\D_{\rm IT}$, its existence,
however, suggests that metastable states exist that should be described
with RSB. We therefore consider the 1RSB ansatz. We divide the $n$ replicas
$a=1,\ldots, n$ into $n/s$ blocks, 
where $s$ is the so-called Parisi parameter \cite{MPV87}. The overlap
matrix becomes
\beq
q_{ab} =\begin{cases}
q_d & a=b\\
q_1 & a,b \textrm{ in the same block}\\
q_0 & a,b \textrm{ in different blocks}
\end{cases}
\eeq
and analogous for $\hat q$. 
For $s$ strictly equal to one we get back the replica
symmetric ansatz Eq.~(\ref{eq:RS:ansatz}).
Note that for $s \neq 1$, $m$ and $q_0$ are in general different in the solution:
this is crucial when evaluating the MSE Eq.~(\ref{eq:MSE})
as the minimum of the MSE does not correspond in general to the
maximum of $m$.

The 1RSB free energy takes the form
\beq
\begin{split}
  \mathrm f_{\text{1RSB}} (\D,s) = {\bf extr} \bigg\{  \phi_{\text{1RSB}} \left( m, q_0, q_1,  \D, s \right) 
 \bigg\}\, ,
\end{split}
\eeq 
with
\beq
\begin{split}
\phi_{\text{1RSB}} \left( m, q_0, q_1,  \D, s \right) 
= 
 \frac{m^2}{2 \D}  - s \frac{q_0^2}{4 \D}  - (1-s) \frac{q_1^2}{4 \D} + \\
- \frac{1}{s} \mathbb{E}_{x^{(0)},W} \left[ 
f \left( \frac{q_1}{\D} , \frac{m}{\D} x^{(0)} + \sqrt{\frac{q_0}{\D}} W , \frac{q_1-q_0}{\D}  \right) 
\right]\, ,
 \label{eq:1RSB_FE}
\end{split}
\eeq 
where
\beq
\begin{split}
 f \left( A,B, C \right) =
 \ln  \int dh \, \sqrt{\frac{C}{2 \pi}} \, e^{-\frac{1}{2} C h^2}  \cdot 
\\
\cdot \left[ \int dx \, P_X (x) \, e^{- \frac{1}{2} A x^2 + \left( B +
      C h \right) x}    \right]^s\, .
\end{split}
\eeq 
The stationary points of the 1RSB free energy are now obtained by
the fixed points of 
\beq
\begin{split}
 m &= \frac{1}{s} \mathbb{E}_{x^{(0)},W} \left[ x^{(0)} 
 \frac{\partial f}{\partial B}  \right]
 \\
 q_0 &= \frac{1}{s^2} \mathbb{E}_{x^{(0)},W} \left[ 
\left( \frac{\partial f}{\partial B} \right)^2  \right]
 \\
 q_1 &= \frac{2}{s (s-1)} \mathbb{E}_{x^{(0)},W} \left[ 
 \frac{\partial f}{\partial A} + \frac{\partial f}{\partial C}   \right]
\end{split}
\label{eq:1RSB_saddle_point}
\eeq 
where
$A= q_1/\D$, $B=m x^{(0)}/\D + W\sqrt{q_0/\D}$ and $C=(q_1-q_0)/\D$ 
and the extremum is a minimum in $m$ and a maximum in the other parameters. 

We would like to reiterate here the observation that in the same
way that the stationary points of the RS free energy correspond to
state evolution fixed points of the AMP algorithm, the stationary
points of the 1RSB free energy correspond to the fixed points of the
state evolution of an approximate survey propagation algorithm that depends on
$s$ \cite{AKUZ18}.
In particular, the expression (\ref{eq:MSE}) exactly gives the MSE of such algorithm with $m$ and $q_0$ being the solution of (\ref{eq:1RSB_saddle_point}).

For high enough $\D$ the 1RSB solution collapses to the RS one,
meaning that $q_0=q_1=m=0$. 
At  $\D_{\rm dyn}$ the saddle point
equations for $s = 1$ admit a solution with $m=q_0=0$,  $q_1>0$.
The value of $q_1$ in this
solution coincides with the value of $m$ in the high magnetization
RS branch discussed in the previous section. At  $\D_{\textrm{IT}}$ the metastable states undergo an entropy 
crisis transition. Although the thermodynamically dominant state
becomes the state with high correlation with the ground truth signal, glassy states 
continue to exist. In fact as far as these states are concerned - if we
neglect the high magnetization state - the system undergoes 
there a Kauzmann transition where the dominant glassy states have zero
complexity and a value of the Parisi parameter $s$ is 
determined by the condition that complexity $\Sigma(\D,s)$ (defined
below) is equal to zero\footnote{Notice the analogy of the high-magnetization state 
here with the crystal state in the physics of glasses.}. 

Let us now discuss $s\ne 1$ solutions. It is well known that the
Parisi parameter $s$ can be
interpreted as an effective temperature that enables to select families of metastable states of given (internal) free energy \cite{Mo95}. Their corresponding complexity $\Sigma$ (defined as the log of their 
number) 
is obtained by deriving
(\ref{eq:1RSB_FE}) w.r.t $s$ \cite{Mo95}, and multiplying the result by $s^2$, i.e.
\beq
\begin{split}
&\Sigma \left(  \D, s \right) = 
\frac{s^2}{4 \D} \left( q_1^2 - q_0^2 \right)  \\
&- s^2 \frac{\partial}{\partial s} \mathbb{E}_{x^{(0)},W} \left[ 
 \frac{1}{s} f  \left( \frac{q_1}{\D} , \frac{m}{\D} x^{(0)} + \sqrt{\frac{q_0}{\D}} W , \frac{q_1-q_0}{\D}  \right) 
\right]
 \label{eq:complexity}
\end{split}
\eeq 
As expected this complexity for $s=1$ coincides with the free energy difference
between the two RS branches discussed in the previous section. 

\begin{figure}[t]
\includegraphics[width=\columnwidth]{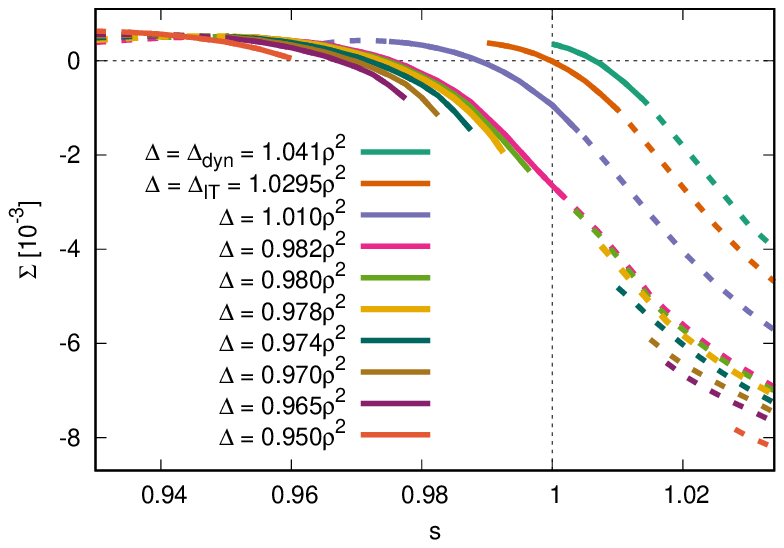}
\vspace{-0.5cm}
\includegraphics[width=\columnwidth]{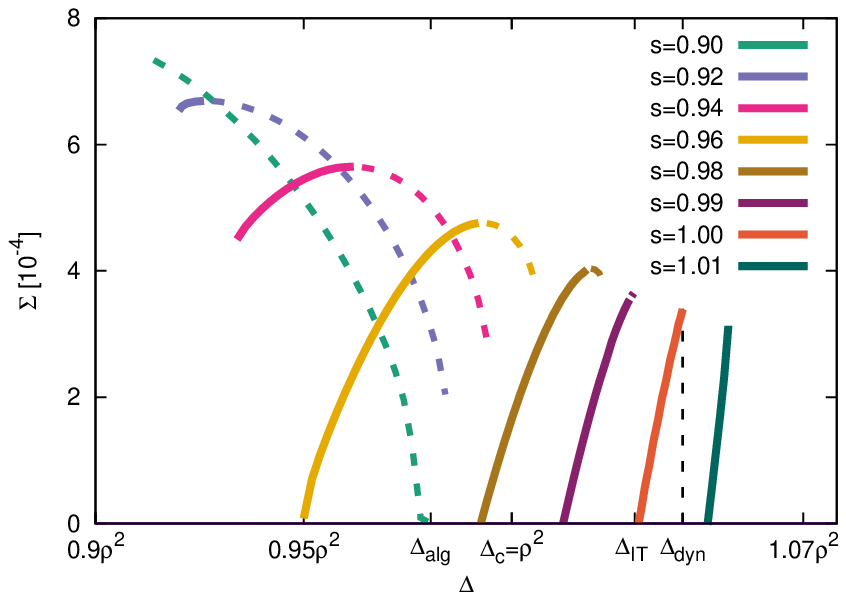}
\caption{The complexity of metastable states $\Sigma$ as a function of the
  Parisi parameter $s$ and the noise $\Delta$, for prior (\ref{eq:RadBern}) with
  sparsity $\rho=0.08$. 
Upper panel, complexity at fixed $s$ in the whole domain of existence of a
non-trivial fixed point.  
Lower panel, the physical region of positive $\Sigma$ as function of $\D$. 
We draw the stable solutions with a solid line and the unstable, wrt the eigenvalues (\ref{eq:eigenvalues}),
with a dashed line.
For each value of $\D\in [\D_{\rm IT},\D_{\rm dyn}]$ the value of 
$\Sigma(\D,s=1)$ represents the complexity of the family of  thermodynamically dominating states. 
Below $\D_{\rm IT}$ the $s=1$ solution in non-physical and
$\Sigma(\D,s=1)<0$.  The algorithmic threshold of AMP occurs when the
ghost-glassy states at $s=1$ have a spinodal transition towards the signal. 
}
\label{complexity}
\end{figure}

In Fig.~\ref{complexity} we plot the complexity as function of both $s$ and of the noise variance $\Delta$.
For each value of $s$ we find two regions: a physical region where $\Sigma$ is positive, and an non-physical one where $\Sigma<0$. 
Note as the physical region with positive complexity continues not
only below $\Delta_{\rm IT}$, but even well below $\Delta_{\rm alg}$.

The 1RSB solution is not guaranteed to give the exact description of the glassy states. 
It is well known that in the replica solutions should be stable against (further) breaking of the replica symmetry.
 This requires that all the eigenvalues of the Hessian of the free energy should be positive in the solution.  
The 1RSB solutions can loose stability in two possible ways, 
associated, to negative values of the following eigenvalues \cite{Ga85, GKS85,MPR04}: 
\beq
\begin{split}
\l_{\text{I}} &= 1-\frac{1}{\D}\int_{-\infty}^\infty \de h P(s,h)\left(f''(s,h)\right)^2\\
\l_{\text{II}} &= 1-\frac{1}{\D}\int_{-\infty}^\infty \de h P(1,h)\left(f''(1,h)\right)^2\:.
\end{split}
\label{eq:eigenvalues}
\eeq
where ($A = q_1/\D$, $B= \frac{m}{\D} x^{(0)} + h$ and $C = (q_1-q_0)/\D$ )
\beq
\begin{split}
 f(1,h) =& \ln \int dx P(x) \exp \left[ -\frac{ A}{2  } x^2 + h x
 \right]\, ,
\\
 f(s,h) =& \frac{1}{s} \ln \int \frac{dz}{\sqrt{2 \pi C}}  e^{-\frac{
     z^2}{2 C}} e^{s f(1,h-z)}\, ,
\end{split}
\eeq
\beq
\begin{split}
 P(s,h) =& \mathbb{E}_{x^{(0)}} \left[  \sqrt{ \frac{\D}{2 \pi  q_0} }
   \exp \left( - \frac{\D}{2 q_0} B^2  \right) \right]\, ,
 \\
  P(1,h) =& e^{s f(1,h)} \int \frac{dz \, e^{-\frac{ z^2}{2 C}} }{\sqrt{2 \pi C}}   
 \cdot P(s,h-z) \, e^{- s f(s,h-z)}\, .
\end{split}
\eeq
A negative $\lambda_{\text{I}}$ (type I instability) signals 
the appearance of new scales of distance between states.  
A negative $\lambda_{\text{II}}$ on the other hand is met when the glassy states are unstable 
against a {\it Gardner transition} to further RSB \cite{Ga85,GKS85}: 
each metastable state splits into a hierarchy of new states (type II instability) \cite{MPR04}. 
In Fig.~\ref{complexity} we mark with full lines the stable region, with dashed lines the unstable ones. 
Type I instability is found for large $s$ in the non-physical region of negative complexity.   
Type II instability is found in the physical region at small values of $s$ and it has been found also in spin glass models \cite{MR03,MPR04,CLR05}.

Let's now discuss in detail the glassy solutions that one finds for 
 $\D<\D_{\rm IT}$ representing metastable states with higher free energy 
than the high-magnetization solution. These solutions  have zero or low magnetization (overlap with the signal).
As already remarked, for a given~$\Delta$, among all the glassy states the ones with lowest
total free energy 
turn out to be the
ones with zero complexity $\Sigma$. For different fixed values of the parameter $s$, the complexity curves reach zero
value at different values of $\D$. Remarkably, as illustrated in Fig.~\ref{complexity} a stable (towards higher
levels of RSB) zero-complexity solution is found down to a value of
noise $\D_{\rm 1RSB,equil} < \D_{\rm alg} $. Stable solutions of positive complexity
exists down to $\D_{\rm 1RSB,stable}< \D_{\rm 1RSB,equil} $, and solutions with
positive complexity (irrespective of the stability) down to $\D_{\rm
  1RSB, all}< \D_{\rm 1RSB,stable}$. Example of specific values for
$\rho=0.08$ in Fig.~\ref{complexity} are $\D_{\rm alg} \sim 0.9805 \rho^2$, $\D_{\rm
  1RSB,equil}  \sim 0.951 \rho^2$, $\D_{\rm 1RSB,stable} \sim 0.918 \rho^2$, $\D_{\rm 1RSB,all} \sim 0.903 \rho^2$. 
This notably means that for $\D<\D_{\textrm{alg}}$, namely in the easy phase where
AMP converges close to the signal,
 families of metastable states continue to exist, some of them being
 stable with
 extensive complexity.

One can discuss how do these states influence
Monte-Carlo dynamics, that explore the space of configuration
according to principles of physical dynamics. On the one hand, one could conjecture 
that Monte-Carlo dynamics gets trapped by glassy states even below 
$\D_{\textrm{alg}}$. On the other hand, the dynamics is expected to fall out of
equilibrium for all $\D< \D_{\rm dyn}$ and it is not a priori clear in
which states it should get trapped. While AMP clearly works for
$\D<\D_{\rm alg}$ and does not work for $\D>\D_{\rm alg}$, our
analysis does not provide any reason why the threshold $\D_{\rm alg}$
should be relevant for Monte Carlo or other sampling-based algorithms.
For such physical dynamics, numerical simulations and analytic studies
in suitable models are necessary to clarify the question of what is
the corresponding algorithmic threshold.

\begin{figure}[t]
\includegraphics[width=\columnwidth]{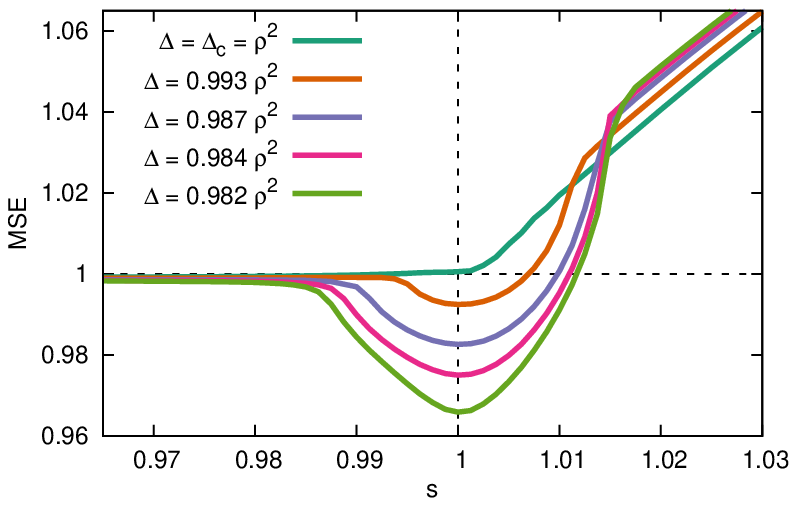}
\includegraphics[width=\columnwidth]{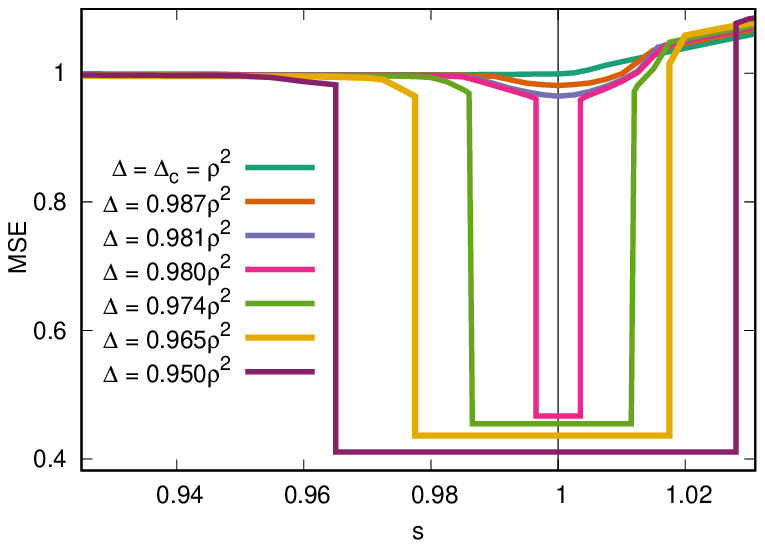}
\caption{
The MSE as a function of the Parisi parameter $s$
  for different values of the noise strength $\D$. 
  The smallest MSE is always reached for  $s=1$, corresponding to the
  performance of the AMP algorithm, with a threshold at
  $\D_{\rm alg}=0.9805\rho^2$.
}
\label{fig:MSE}
\end{figure}

So far we focused on glassy states of positive complexity
(i.e. existing with probability one for typical instance). There are also solutions of the
1RSB equations having negative complexity. We will call the negative-complexity solution the {\it
  ghost-glassy} states. From the physics point
of view those solutions do not correspond to physical states for typical
instances. Yet, from the algorithmic point of view they do correspond
to the fixed points of the ASP algorithm \cite{AKUZ18} run for a given value of
Parisi parameter  $s$, as such they can be reached algorithmically. 
At this point it becomes relevant to understand for which value
 $\D_{\rm alg}(s)$ do the ghost-glassy
state disappear, developing a spinodal instability towards the
high-magnetization state.  In particular we can ask the natural question if with a
suitable choice of the Parisi parameter $s$ the ASP improves over
the algorithmic threshold $\Delta_{\rm alg}\equiv \D_{\rm alg}(s=1)$
of the usual AMP ($s=1$) and if we could have an $s$ for which 
$\D_{\rm alg}(s)>\D_{\rm alg}(1)$. With this question in mind in Fig.~\ref{fig:MSE} we plot
the mean-squared error (MSE) with the ground truth signal given by Eq.~(\ref{eq:MSE}) as a
function of $s$ for various values of $\D$. We initialize the 1RSB
fixed point equations at infinitesimal magnetization and iterate them
till a fixed point. We observe that for all values of $\D$ the MSE is
minimized for $s=1$, i.e. by the canonical AMP algorithm.

\section{Conclusion}
\label{sec:conclusion}

In conclusion, we studied the glassy nature of the hard phase in inference problems. Our results
imply that indeed the corresponding metastable state is glassy, i.e. composed of
exponentially many states. We evaluate their number (complexity) as a
function of their internal free energy to conclude that this
glassiness extends to a range of the noise parameter $\D$ even larger
than the extent of the the hard phase. This finding re-opens the
natural question of performance limits of Monte-Carlo based
sampling. While some recent works \cite{decelle2011asymptotic} anticipated numerically that
Monte-Carlo and message passing will share the same algorithmic threshold,
our results do not provide any evidence of this. 
Instead they suggest that since glassiness is present
  also below the algorithmic threshold of AMP the performance of
  sampling-based algorithms will be different in
  general.  In order to validate this proposition one
    needs to study a different model than the
    present one. The present model is dense and thus not suitable for
    large scale simulations, also analytically tractable description of
    sampling-based dynamics for the present model is a major open
    question. One possibility is to perform large-scale numerical
    study with Monte-Carlo based dynamics in
    diluted models such as those studied in \cite{ricci2018typology}. Another
    possibility is to aim at analytical description of the Langevin
    dynamics that is known in a tractable form so far only for mixtures of spherical $p$-spin models.
 
 While we anticipate that the performance of the usual sampling-based algorithms will
be hampered by the glassiness, it is an interesting open question to
investigate whether other algorithms are able to match the performance
of AMP. We have in mind for instance the algorithms based on the robust
ensemble as introduced in \cite{baldassi2016unreasonable}.

Concerning the AMP algorithm,  we conclude that, 
despite the fact that it assumes the hard-phase not to be glassy, 
the improved description in terms of one-step replica symmetry breaking, that takes glassiness into account, does not 
provide algorithmic improvement. This is at variance with the
situation in random constraint satisfaction problems, where the
knowledge  of the organization space of solutions provided by 1RSB leads to
algorithmic improvement \cite{BMZ05}. We note that this
  observation is surprising, and we are missing 
a physically intuitive explanation for why taking glassiness into
account improves performance in optimization problems but not in
Bayes-optimal inference. 
We stress that our results provide strong evidence towards
the conjecture that the hard phase is impenetrable for some computationally
fundamental reasons. Further investigation of this is an exciting
direction both for physics and theoretical computer science.

In this paper we use the
example of low-rank matrix estimation with spins $0$ and $\pm 1$ as a
prototypical example in which the hard phase exists. We checked that
the resulting picture applies in a range of parameters and also for
some other models (such as planted mixed $p$-spin model) where the hard phase
was identified. We expect the picture presented here to be generic in
all the problems where the hard phase related to a first order phase
transition was identified.

We also note that our above conclusions apply to the case of
Bayes-optimal inference where the generative model is matched to the
inference model. In case the hyper-parameters are not known or
mismatched the message passing algorithm that takes glassiness into
account can provide better error and robustness, this is investigated
in detail in \cite{AKUZ18}.

Finally, we mention that the results shown here may be compelling also beyond inference problems.
In particular, the instabilities 
of the RS solution at $\D_{\rm alg}$ and $\Delta_{\textrm c}$  can be related to a  similar 
phenomenon occurring in the mean field theory of liquids and glasses \cite{PZ10, CKPUZ17}. 
A phase structure similar to the one presented 
in this paper is found in that case, if we identify $\D$ as analogue to 
an (inverse) density parameter and the reconstruction phase as the crystal. Also in that case, the RS solution
representing the liquid at low density describes a non-ergodic extensive complexity phase at higher density. As it is the case here, 
there is a density where  complexity vanishes, but the solution can be continued below this point.  Finally, there is a 
maximum density where the solution undergoes an instability - called Kirkwood instability - and ceases to exist \cite{FP99, MK11}. 
Our analysis  suggests that within inference models not only the non-physical negative complexity RS solution could undergo this instability, but also the glassy ones. 
Whether this phenomenon could be relevant for other glassy systems is an intriguing question.

\section*{Acknowledgments}
We would like to thank Giulio Biroli, Florent Krzakala, and Guilhem
Semerjian for fruitful discussions. 
This work is supported by \textquotedbl Investissements d'Avenir\textquotedbl  \ LabEx PALM (ANR-10-LABX-0039-PALM) (SaMURai and StatPhysDisSys projects), and from the ERC under the European
Unions Horizon 2020 Research and Innovation Programme Grant
Agreement 714608-SMiLe. The  work of SF was supported by a grant from the 
Simons Foundation (No. 454941, Silvio Franz).

\bibliography{refs_LowRank}

\end{document}